\documentclass[12pt,pra,onecolumn,showpacs]{revtex4}

\begin{document}

\title{Atom-Pair Tunneling and Quantum Phase Transition in Strong
Interaction Regime }
\author{J.-Q. Liang}
\email{jqliang@sxu.edu.cn}
\affiliation{Institute of Theoretical Physics and Department of Physics, Shanxi
University, Taiyuan, 030006, China}
\author{J.-L. Liu}
\affiliation{Institute of Theoretical Physics and Department of Physics, Shanxi
University, Taiyuan, 030006, China}
\author{W.-D. Li}
\affiliation{Institute of Theoretical Physics and Department of Physics, Shanxi
University, Taiyuan, 030006, China}
\author{Z.-J. Li}
\email{zjli@sxu.edu.cn}
\affiliation{Institute of Theoretical Physics and Department of Physics, Shanxi
University, Taiyuan, 030006, China}

\begin{abstract}
We propose a Hamiltonian of ultracold spinless atoms in optical lattices
including the two-body interaction of nearest neighbors, which reduces to
the Bose-Hubbard model in weak interaction limit. An atom-pair hoping term
appearing in the new Hamiltonian explains naturally the recent experimental
observation of correlated tunneling in a double-well trap with strong
atom-atom interactions and moreover leads to a new dynamic process of
atom-pair tunneling where strongly interacting atoms can tunnel back and
forth as a fragmented pair. Finally a new dynamics of oscillations induced
by the atom-pair tunneling is found in the strong interaction regime, where
the Bose-Hubbard model gives rise to the insulator state with fixed
time-averaged value of atom-occupation-number only. Quantum phase
transitions between two quantum phases characterized by degenerate and
non-degenerate ground states are shown to be coinciding with the Landau
second-order phase transition theory. In the system of finite atom-number
the degeneracy of ground states can be removed by quantum tunneling for the
even-number of atoms but not for the odd-number.
\end{abstract}

\maketitle

\section{Introduction}

The ultracold atom-gas clouds possess many advantages for
investigation of quantum phenomena and hence become a test ground of
quantum mechanical principles in many extraordinary aspects and new
prospective regime. Recently rapid advances of experimental
techniques in optical traps open up a prospect for the study of
quantum phase transition (QPT), for example, from a superfluid to a
Mott-insulator\cite{greiner,spielman,stoferle}, where the ratio
between tunnel coupling through the interwell barriers and the
atom-atom interaction plays a crucial role. Up to date the QPT has
been studied only based on the well known and widely applied
Bose-Hubbard (BH) Hamiltonian, which in a mean field description
leads to coupled equations of population-imbalance and phase
difference between two wells for a double-well trap and thus can
describe the Josephson oscillation as well as self-trapping of
Bose-Einstein condensates (BECs) qualitatively. It is demonstrated
that the investigation of experimentally observed non-linear
self-trapping\cite{albiez} could provide a test ground of the mean
field description of the BH model in the strong nonlinear regime.
Particle tunneling through a classically prohibiting potential
barrier is one of the characteristic effects of quantum physics,
which has been essentially studied in a single-particle manner. For
the many-body case strong interactions between particles may
fundamentally alter the tunnel configuration and result in a
correlated tunneling, which was explored most recently in ultracold
atoms\cite{folling,zollner}. On the other hand a co-tunneling regime
can be achieved in coupled mesoscopic quantum dots, where separate
electrons only tunnel in a correlated way\cite{de,zum}. We point out
in this paper that the well known BH Hamiltonian based on the
hard-core interaction is not able to describe the dynamics of
atom-pair tunneling, which is the dominant dynamic effect in strong
interaction regime, since the correlated atom-pair tunneling
requires long range correlation of wave functions. The BH
Hamiltonian, which is valid in a relatively weak interaction regime,
should be extended in the strong interacting regime to include the
superexchange interactions between atoms on neighboring lattice
sites\cite{duan,kuklov}. A peculiar atom-pair hopping term appearing
in the new Hamiltonian explains very well the recently reported
experimental observation of correlated tunneling and moreover leads
to new dynamics in superstrong interaction regime, which has not yet
been explored.

The Hamiltonian of second-quantization beyond the on-site approximation is
seen to be (see appendix for derivation)
\begin{eqnarray}
H &=&\sum\limits_{i}\varepsilon
_{i}n_{i}-\sum_{i}[J_{i}-U_{3}(n_{i}+n_{i+1}-1)](a_{i}^{\dag
}a_{i+1}+a_{i+1}^{\dag }a_{i})+\frac{U_{0}}{2}\sum\limits_{i}n_{i}(n_{i}-1)
\nonumber \\
&&+(U_{1}+U_{2})\sum\limits_{i}n_{i}n_{i+1}+\frac{U_{2}}{2}%
\sum\limits_{i}(a_{i}^{\dag }a_{i}^{\dag }a_{i+1}a_{i+1}+a_{i+1}^{\dag
}a_{i+1}^{\dag }a_{i}a_{i})  \label{1}
\end{eqnarray}%
where $a_{i}$ is the $i$th-site boson annihilation operator and $%
n_{i}=a_{i}^{\dag }a_{i}$ is the corresponding atom-number operator. The new
coupling constants between atoms on neighboring lattice-sites are given by
\[
U_{1}=\int
w_{i}(x_{2})w_{i+1}(x_{1})U(|x_{1}-x_{2}|)w_{i+1}(x_{1})w_{i}(x_{2})dx_{1}dx_{2},
\]%
\[
U_{2}=\int
w_{i}(x_{2})w_{i}(x_{1})U(|x_{1}-x_{2}|)w_{i+1}(x_{1})w_{i+1}(x_{2})dx_{1}dx_{2},
\]%
and
\[
U_{3}=\int
w_{i}(x_{2})w_{i+1}(x_{1})U(|x_{1}-x_{2}|)w_{i+1}(x_{1})w_{i+1}(x_{2})dx_{1}dx_{2},
\]%
where $U(|x_{1}-x_{2}|)$ is the two-atom interaction potential and $w_{i}$
denotes Wannier wave functions. $U_{0}$ is the usual on-site
atom-interaction strength, and the fourth term relating to $U_{1}$ denotes
the nearest-neighbor repulsion, which has been given in various lattice
models in the literature, while the $U_{2}$ part is new. The conceptually
new transition-matrix element $U_{2}/2$\ describes obviously the atom-pair
tunneling. The atom-atom interaction including the nearest-neighbor results
in an additional Josephson tunneling term in relation with the coupling
constant $U_{3}$ and the atom-number operator $n_{i}+n_{i+1}-1$. This term
as a matter of fact suppresses the interwell hopping. The coupling constants
$U_{1}$, $U_{2}$ , and $U_{3}$ are much smaller comparing with $U_{0}$ and
can be evaluated roughly with Gaussian Wannier function and $\delta $%
-function potential. We have
\[
U_{1}\approx U_{2}\approx \epsilon ^{2}U_{0},\qquad U_{3}\approx \epsilon ^{%
\frac{3}{2}}U_{0},
\]%
where
\[
\epsilon =e^{-\frac{\pi ^{2}s^{1/2}}{4}},\qquad s=\frac{V}{E_{R}},
\]%
with $V$ , $E_{R}$ being the well depth and lattice recoil energy
respectively. Using the experimental values\cite{folling} for $V$ and $%
E_{R} $ we find from the equations above that $U_{1}\approx U_{2}\approx
0.02U_{0}$. This is in close agreement with the values we take later on with
best fitting of the curves with experimental data points $%
U_{1}=U_{2}=0.018U_{0}$ (see below).

When a correlated many-body quantum-system is driven by a
controllable parameter, the ground-state energy may have a
structural change at a critical value of this parameter. This
phenomenon is called QPT from a disordered phase to an ordered one
obeying the Landau theory with ordinary symmetry breaking. A
well-studied example for the QPT is a model described by the
Hamiltonian of Lipkin-Meshkov-Glick (LMG) type, which is originally
introduced in nuclear physics as exactly solvable many-body
interacting model and now has been used in a broad range of topics
such as spin system and BECs. In the spin system QPT between
ferromagnetic phase of long-range magnetic order and disordered
paramagnetic phase can occur at a critical value of external
magnetic filed. We show that the atom-pair tunneling term results in
new dynamics of BEC in a double-well trap and a formally same QPT as
that in the spin system can be found in the strong atom-atom
interaction regime, where the BH Hamiltonian could only give rise to
an insulator-phase with a fixed occupation number of atoms in each
well. In this paper a new method is adopted for the investigation of
QPT, where the many-body system is converted into a giant
pseudo-spin which in turn is mapped to a single particle problem
with an effective potential. Moreover for the finite-size system
with a small number of particles the quantum tunneling may play a
significant role and as a consequence the QPT would depend on the
number-parity (even or odd) of particles resulted from the
topological phase interference of tunneling paths. This peculiar
phenomenon is analyzed in section IV.

\section{ Two-atom dynamics in a double-well trap and experimental evidence
of atom-pair tunneling}

The tunneling dynamics of a few atoms loaded in a double-well trap has been
studied by varying the interaction strength from weak to strong limit and it
is shown for the two-atom case that the character of tunneling changes from
Rabi oscillation to a correlated process with increasing interaction\cite%
{zollner,folling}, namely, when repulsive interactions are strong,
two atoms located on one side of the barrier cannot separate, but
tunnel together as a pair in a co-tunneling process. As a matter of
fact interactions of ultracold atoms can be adjusted experimentally
over a wide range via Feshbach resonances which make it possible to
explore the limit of strong correlation. A direct observation of the
correlated tunneling was reported recently\cite{folling} and
theoretical analysis has been also presented in terms of two-body
quantum mechanics\cite{zollner}. Moreover it has been demonstrated
that\ if the bosons repel each other infinitely strongly, they can
be mapped to noninteracting fermions (the fermionization limit) in
the sense that the hard-core interaction mimics the exclusion
principle. Near fermionization the strongly interacting atoms tunnel
back and forth as a fragmented pair in a double-well trap
\cite{zollner}.

For the double-well potential the Hamiltonian Eq.(1) reduces (with a
constant-energy term renormalized to zero) to
\begin{eqnarray}
H &=&\frac{\Delta }{2}(n_{L}-n_{R})-J(a_{L}^{\dag }a_{R}+a_{R}^{\dag }a_{L})+%
\frac{U_{0}}{2}\left[ n_{L}(n_{L}-1)+n_{R}(n_{R}-1)\right]  \nonumber \\
&&+(U_{1}+U_{2})n_{L}n_{R}+\frac{U_{2}}{2}(a_{L}^{\dag }a_{L}^{\dag
}a_{R}a_{R}+a_{R}^{\dag }a_{R}^{\dag }a_{L}a_{L}),  \label{2}
\end{eqnarray}%
where $\Delta =\varepsilon _{L}-\varepsilon _{R}$ is the bias potential
between two wells and
\[
J=J_{0}-(N-1)U_{3}
\]%
with $J_{0}$ being the single-atom Josephson coupling constant. The
effective Josephson coupling constant $J$ , which is suppressed by the
nearest-neighbor repulsion coupling, can become a negative value for
sufficiently large number of atoms $N$. In tight-confinement approximation,
the quantum state of atoms can be described in a Fock state basis $%
|n_{L},n_{R}>$ with $n_{L}$ and $n_{R}$ being non-negative integers. For
one-atom occupation in the two wells corresponding to a filling factor $1/2$
of the optical lattice, the Hamiltonian Eq.(2) gives rise to nothing new but
the Josephson oscillations (see below). We are interested in the dynamics of
two-atom occupation (filling factor $1$ ), where the Fock-state base-vectors
are $|1,1>,|2,0>$ and $|0,2>$. The states $|0,2>,|2,0>$ both couple to the
state $|1,1>$ by the Josephson-tunnel matrix-element $J$ and couple each
other directly via the matrix element $U_{2}$, the effect of which is
vanishingly small in weak interaction regime (roughly speaking $U_{0}<J$ ),
where the Josephson tunneling dominates. On the other hand for strong
interactions ($U_{0}>J$) the energy difference between states $|0,2>$ and $%
|1,1>$ is larger than the coupling $J$ and therefore the Josephson-tunneling
induced transition between these states is suppressed due to the strong
detuning. However the direct transition between states $|0,2>,$ and $|2,0>$
via $U_{2}$ matrix element is still resonant and gradually becomes the
dominant dynamics with continuously increasing interactions. The BH
Hamiltonian can only give rise to the first-order transitions between states
$|1,1>$ and $|0,2>$ (or $|2,0>$), while the transition, for example, from
state $|2,0>$ to state $|0,2>$ is generated only by a second-order transition%
\cite{folling} via the $J$ matrix element that $|2,0>\rightarrow
|1,1>\rightarrow |0,2>$ with an effective coupling constant $2J^{2}/U_{0}$,
which alone, we will see, is too small to be responsible for the correlated
tunneling. We demonstrate in this paper that the correlated tunneling in
fact includes both the second-order transition and the atom-pair tunneling
via the matrix element $U_{2}$, which is more important dynamic process in
the superstrong interaction regime ($U_{0}>>J$) near fermionization. To see
this closely we can evaluate the energy eigenvalues $E_{i}$ ($i=1,2,3$) and
the corresponding eigenstates $|\psi _{i}>$ by direct diagonalization. The
two-atom dynamics is studied by solving the time-dependent Schr$\ddot{o}$%
dinger equation $i\hbar \frac{d}{dt}|\psi (t)>=H|\psi (t)>$ within a
three-state description for both the Eq.(2) and BH Hamiltonians. Thus the
time-evolution of average position of the two atoms can be evaluated by
\begin{equation}
<x>=\frac{1}{2}<\psi (t)|n_{L}-n_{R}|\psi (t)>,
\end{equation}%
which characterizes the tunneling dynamics of atoms. We consider the two
atoms initially localized on one side of the double wells, for example the
left-well. The time-evolutions of average positions are shown in Fig.1
(solid lines) and are compared with the measured time-resolved traces (black
dots). If the interaction energy $U_{0}$ is much smaller than the tunnel
coupling constant $J$ ($J/U_{0}=1.5$ in the experiment \cite{folling}), the
Josephson tunneling is a dominant dynamic mechanism and the average-position
oscillation contains more than one frequency component indicating the
transition between the two states $|2,0>$ and $|0,2>$, which is induced
mainly by the second-order process of the Josephson-matrix element \cite%
{folling}. The direct transition between the two states $|2,0>$ and
$|0,2>$ via matrix element $U_{2}$, which is more than two-order
lower than the second-order Josephson tunneling, is negligibly small
and thus results essentially in no effect seen from Fig.1(a).
However, when the interaction
energy $U_{0}$ increases to reach the interaction-dominated regime ($%
J/U_{0}=0.2$ in the experiment \cite{folling}), where the Josephson
tunneling is suppressed due to the energy difference between states
$|0,2>$ (or $|2,0>$) and $|1,1>$, the main dynamic process is the
correlated tunneling with frequency $550Hz$ (oscillation period
1.8$ms$) seen from the measured two-atom average position
\cite{folling} (Fig.1(b) black dots), which coincides obviously with
the numerical result of the Hamiltonian Eq.(2) (red-solid line in
Fig.1(b)) including direct pair-tunneling process between states
$|0,2>$ and $|2,0>$ via the transition element $U_{2},$ which leads
to a significant modification of the dynamics. The interaction
parameters $U_{1}=U_{2}=0.018U_{0}$ are determined by the best
fitting with the experimental data \cite{folling} for a fixed
Josephson coupling constant $J_{0}$, which is obtained from the
experiment of single-atom occupation \cite{folling} (see below). We
see that the Josephson tunneling between the states $|1,1>$ and
$|0,2>$ (or $|2,0>$) is visible as a small modulation with a period
of $400\mu s$ . The second-order transition $|0,2>\rightarrow
|1,1>\rightarrow |2,0>$ (or $|2,0>\rightarrow |1,1>\rightarrow
|0,2>$) with
the coupling constant $2J^{2}/U_{0}$, which is more than four times of the $%
U_{2}$ value in the present case, is indeed the dominant dynamic process as
demonstrated in Ref.\cite{folling}. It was, however, realized in Ref.\cite%
{folling} that the theoretical results evaluated from the BH
Hamiltonian with the second-order transition give rise to slightly
longer oscillation-period of average position and lower oscillation
amplitude (Fig.1(b), blue-dash line) than the experimental data
(black dots\cite{folling}). This deviation although small is crucial
and was compensated by modifying the Josephson coupling constant
with an additional 3-10\% higher value in Ref.\cite{folling}. We
show that this problem can be cured by the atom-pair hoping via the
transition element $U_{2}$. When the interaction strength increases
up to the value that $J/U_{0}=0.1$, the atom-pair hoping becomes
more important since the coupling constant of second-order
transition is about the same value of $U_{2}$ and thus the BH
Hamiltonian fails to describe tunneling dynamics. The average
positions of the two atoms evaluated from Eq.(2) (red-solid line)
and BH Hamiltonian (blue-dash line) are given in Fig.1(c) showing a
great difference, that the oscillation frequency from Eq.(2) is
almost two times higher than that evaluated from the BH Hamiltonian
for the transition between states $|0,2>$ and $|2,0>$ (i. e. the
oscillation of large amplitude in Fig.1(c)). Increasing the
atom-atom interaction the atom-pair hopping introduced in \ this
paper becomes more and more important than the Josephson tunneling
and at a critical point $J=J_{0}-(N-1)U_{3}=0$, where the Josephson
tunneling
vanishes, we have the sinusoidal population oscillations between two states $%
|2,0>$, $|0,2>$ induced only by the atom-pair tunneling i.e. the effective
Josephson oscillation of atom-pair. We see that atom-pair tunneling becomes
the dominant dynamic process and should be observed experimentally.

In addition to the average center-of mass positions the phase relation
between two wells is also measured experimentally by a separate
interferometric sequence of double-slit matter-wave interference pattern,
which can be evaluated by
\begin{equation}
P=\sum_{i,j}<a_{i}a_{j}^{+}>e^{i\mathbf{k\cdot }(\mathbf{r}_{i}-\mathbf{r}%
_{j})}=N(1+\mathcal{V}\cos [\phi +\mathbf{k\cdot }(\mathbf{r}_{L}-\mathbf{r}%
_{R})])  \label{4}
\end{equation}%
where $\phi $ is the relative phase between two wells, $\mathcal{V}$ is the
visibility and $N$ denotes the total number of atoms in the double-well. For
a single-atom occupation the experimentally observed signal is still
sinusoidal however with a correspondingly lower frequency\cite{folling} in
the strong interaction regime ($J/U_{0}=0.2$). Both Hamiltonians give rise
to the same expected sinusoidal population oscillation between two states $%
|1,0>$, $|0,1>$ with oscillation-period depending only on the
Josephson-matrix element $J=J_{0}$. As a comparison the
corresponding average center-of mass position, phase and visibility
are shown in Fig.2 with black dots representing the experimental
data\cite{folling} from which Josephson coupling constant $J_{0}$ is
obtained $J_{0}=0.55\times 10^{3}Hz\cdot $ $h$, where $h$ is the
Planck's constant. We see that the two Hamiltonians make no
difference in description of dynamic behavior of single-atom
occupation seen from the coinciding red (solid) (proposed
Hamiltonian) and blue (dash) (BH Hamiltonian) lines.

\section{Atom-atom interaction induced dynamics and quantum phase transition}

Dynamics of cold atoms in a double-well trap has been well studied based on
the BH Hamiltonian. The $\pi $-phase oscillation, in which the time-averaged
value of the phase difference between two wells is equal to $\pi $, has been
found in the weak interaction limit ($J>>$ $U_{0}$). In the strong
interaction region ($U_{0}>>J$) the BH Hamiltonian results in the
insulator-phase only, where the system is in the fixed atom
occupation-number state. We show that the atom-pair tunneling in the
Hamiltonian Eq.(2) can lead to a new dynamics of atom occupation-number
oscillation, which has not yet been explored. We in this paper provide an
analytic investigation based on an effective Hamiltonian of single-particle
with canonical variables: the atom-number difference (or population
imbalance) and phase difference between the two wells. For the $N$-atom
occupation (filling factor $\frac{N}{2}$ ) in the double-well trap we
introduce the pseudo-angular momentum operators defined by
\begin{equation}
S_{x}=\frac{1}{2}\left( a_{L}^{\dag }a_{R}+a_{R}^{\dag }a_{L}\right) ,
\end{equation}%
\begin{equation}
S_{y}=\frac{1}{2i}\left( a_{L}^{\dag }a_{R}-a_{R}^{\dag }a_{L}\right) ,
\end{equation}%
and
\begin{equation}
S_{z}=\frac{1}{2}\left( n_{L}-n_{R}\right) ,
\end{equation}%
with the total angular momentum
\[
S^{2}=\frac{N}{2}(\frac{N}{2}+1).
\]%
The Hamiltonian Eq.(2) can be written as
\begin{equation}
H=-2JS_{x}-\Delta S_{z}+K_{1}S_{z}^{2}+K_{2}S_{x}^{2},  \label{8}
\end{equation}%
with the parameters given by $K_{1}=U_{0}-U_{1}$, $K_{2}=2U_{2}$, which is
nothing but a LMG model Hamiltonian\cite{lipkin}. Here we adopt a new method
to study the energy spectrum and related QPT of Hamiltonian Eq.(8). To this
end we begin with the Schr$\ddot{o}$dinger equation
\begin{equation}
H\Phi (\phi )=E\Phi (\phi ),  \label{9}
\end{equation}%
where $E$ is the energy eigenvalue to be determined and the generating
function is constructed in terms of the spin function of $S_{z}$
representation such as
\begin{equation}
\Phi (\phi )=\sum_{m=-s}^{s}\frac{C_{m}}{\sqrt{(s-m)!(s+m)!}}e^{im\phi }.
\end{equation}%
in which the pseudo-spin operators have the form of differential operators
\[
S_{x}=s\cos \phi -\sin \phi \frac{d}{d\phi },
\]%
\[
S_{y}=s\sin \phi +\cos \phi \frac{d}{d\phi },
\]%
\[
S_{z}=-i\frac{d}{d\phi }.
\]%
Obviously $S_{z}$ and $\phi $ are two canonical variables describing the
atom-number and phase differences respectively between the double wells. In
the following we may convert our eigenvalue problem Eq.(9) to an effective
single-particle Hamiltonian by making use of a proper unitary transformation
and introducing an incomplete elliptic integral coordinate such that
\begin{equation}
x=\int_{0}^{\phi }\frac{d\phi }{\sqrt{1-\lambda ^{2}\sin ^{2}\phi }}=F(\phi
,\lambda ),
\end{equation}%
where $F(\phi ,\lambda )$ is the incomplete elliptic integral of the first
kind with modulus $\lambda ^{2}=K_{2}/K_{1}$, and then can achieve an
effective single-particle in a quasi-periodic potential $V(x)$ for the
symmetric double-well $\Delta =0$,
\begin{equation}
\left[ -K_{1}\frac{d^{2}}{dx^{2}}+V(x)\right] \psi (x)=E\psi (x),  \label{12}
\end{equation}%
\begin{equation}
V(x)=(a-\lambda ^{2}V_{\min })\frac{\left( cn(x)-\mu \right) ^{2}}{dn^{2}(x)}%
,
\end{equation}%
with
\[
V_{\min }=\frac{a(\xi _{1}-\xi _{2})^{2}-b(\xi _{1}-\xi _{2})+c}{\lambda
^{\prime 2}+\lambda ^{2}(\xi _{1}-\xi _{2})^{2}},
\]%
\[
\xi _{1}=\sqrt{\xi _{2}^{2}+\frac{\lambda ^{\prime 2}}{\lambda ^{2}}},\xi
_{2}=\frac{1}{b}(\frac{a\lambda ^{\prime 2}}{\lambda ^{2}}-c),
\]%
and
\[
\mu =\frac{b}{2(a-\lambda ^{2}V_{\min })},
\]%
where $cn(x)$ and $dn(x)$ denote the Jacobian elliptic functions of modulus $%
\lambda ^{2}$ and $\lambda ^{\prime 2}=1-\lambda ^{2}$. Three parameters $%
a,b,$ and $c$ are related to the model parameters given by
\[
a=K_{2}\frac{N}{2}(\frac{N}{2}+1)-J^{2}/K_{1},
\]%
\[
b=J(N+1),
\]%
and
\[
c=\frac{J^{2}}{K_{1}}.
\]%
The effective-Hamiltonian eigenvalue-problem Eq.(12) corresponds to
a single-particle of mass $m=1/2K_{1}$ in an effective potential
$V(x)$ plotted in Fig.3, which possesses two degenerate minima
located at
\begin{equation}
x_{\pm }=\pm cn^{-1}(\mu )\mathop{\rm mod}[4\mathcal{K}(\lambda ^2)],
\end{equation}%
respectively, where $\mathcal{K(}\lambda ^{2})$ is the complete elliptic
integral of the first-kind with modulus $\lambda ^{2}$. The two degenerate
minima are separated by the central barrier located at $x=0$ with the
barrier height given by
\[
V_{\max }=(a-\lambda ^{2}V_{\min })(1-|\mu |)^{2},
\]%
which can be controlled by the Josephson coupling constant $J$. Particularly
when the Josephson coupling constant vanishes $J=0$, we have $\mu =0$ . the
degenerate minima are located at $x_{\pm }(\mu =0)=\pm \mathcal{K}(\lambda
^{2})\mathop{\rm mod}[2\mathcal{K}(\lambda ^2)]$ and the potential $V(x)$
becomes periodic. Notice the relation between the elliptic-integral
coordinate $x$ and the phase-angle $\phi $ the two degenerate ground states
correspond to the degenerate $\pm \frac{\pi }{2}\mathop{\rm mod}(\pi )$%
-phase states of cold atoms in the double-well trap (Fig.4(a)). The QPT has
obvious meaning in a real spin system, in which $S_{i}$ ($i=x,y.z$) denote
the collective spin-operators. The degenerate $\pm \frac{\pi }{2}%
\mathop{\rm mod}(\pi )$-phase states are nothing but the
ferromagnetic phase of long range magnetic order (gapless) with two
degenerate equilibrium-orientations of the magnetization. When the
Josephson coupling constant $J$ increases the height of center
potential-barrier decreases (see Fig.4) and the two-fold degeneracy
of the ground state gradually lifts. At a critical point (Fig.4(c))
\begin{equation}
\mu _{c}=\pm 1
\end{equation}%
the barriers located at $x_{b1}=0\mathop{\rm
mod}[4\mathcal{K}(\lambda ^2)]$ (corresponding to $\phi =0\mathop{\rm mod}%
(2\pi )$) and $x_{b2}=2\mathcal{K}(\lambda ^{2})%
\mathop{\rm
mod}[4\mathcal{K}(\lambda ^2)]$ (corresponding to $\phi =\pi \mathop{\rm mod}%
(2\pi )$) vanish, and we have the disordered phase called the
paramagnetic phase with a energy gap in the spin language, which in
our case of could atoms in a double-well trap indicates zero-and
$\pi $-phase oscillations. The QPT from the long-range magnetic
order to disordered phases is of the second-order coinciding with
the Landau second-order phase transition theory, where the
dimensionless barrier height
\[
h=\frac{V_{\max }}{(a-\lambda ^{2}V_{\min })}=(1-|\mu |)^{2}
\]%
may be chosen as the order-parameter. The highest value of order-parameter $%
h=1$ ($\mu =0$) corresponds to the degenerate $\pm \frac{\pi }{2}%
\mathop{\rm
mod}(\pi )$-phase states i.e. the ferromagnetic phase of two-fold degeneracy
and the magnetic order decreases with the decreasing order-parameter $h$
(increasing Josephson coupling constant $J$ ). When the order-parameter
vanishes $h_{c}=0$ ($|\mu _{c}|=1$) by properly adjusting the parameters $%
J_{0}$ and $N$, the system approaches the non-degenerate $0\mathop{\rm mod}%
(2\pi )$-phase ( $\mu _{c}=1$) or $\pi \mathop{\rm mod}(2\pi
)$-phase ( $\mu _{c}=-1$) ground state i.e. the paramagnetic phase
in the spin language (Fig.4(c)). The phase-plane portraits
corresponding to the classical Hamiltonian of Eq.(12)
$H=\frac{p^{2}}{2m}$ $+V(x)$ are plotted in the Fig.4 in comparison
with the variation of potential barrier. The closed orbits show the
oscillations with fixed average phase-difference between the
double-well trap and waved-open lines indicate atom-number
self-trapping. The variation of energy gap as a function of
$J/U_{0}$ with various $K_{2}$ is plotted in Fig.5 showing clearly
the smooth second-order phase transition. It is worthwhile to remark
that the entirely new method, which converts the many-body system to
an effective single-particle in a potential field, has obvious
advantage to study the dynamics and QPT of a complex system in a
simple and visible way.

\section{Quantum tunneling and atom-number parity effect}

\bigskip In the previous section the dynamics of cold atoms trapped in a
double-well potential has been investigated in terms of two conjugate
variables, i.e. the atom-number-occupation difference and the relative phase
in the elliptic integral coordinate $x$. In the case of large atom-number $N$
quantum tunneling effect becomes negligibly small, however for small $N$,
the quantum tunneling has to be taken into account. We demonstrate an atom
number-parity effect of tunneling resulted from the boundary condition of
the wave function $\Phi (\phi )$, which is obviously periodic for
even-number of $N$ \ (and therefore the pseudo-spin $s=\frac{N}{2}$ is
integer) and antiperiodic for odd-number
\begin{equation}
\Phi (\phi +2\pi )=\Phi (\phi )e^{i2\pi s},
\end{equation}%
while the boundary condition of wave function $\psi (x)$ in the elliptic
integral coordinate $x$ is
\begin{equation}
\psi (x+4\mathcal{K})=(-1)^{2s}\psi (x).  \label{17}
\end{equation}%
We begin with the transition amplitude, i. e. the Feynman
propagator, between two degenerate vacua $x_{\pm }$ induced by the
quantum tunneling, which can be studied by the instanton technique
\begin{equation}
<x_{+}(T)|x_{-}(-T)>=\int \mathcal{D}\{x\}e^{-S_{E}}
\end{equation}%
where
\begin{equation}
S_{E}=\int_{-T}^{T}\mathcal{L}_{E}d\tau
\end{equation}%
is Euclidian action defined with the imaginary time $\tau =it$ and the
Euclidian Lagrangian is given by
\begin{equation}
\mathcal{L}_{E}=\frac{1}{2}m(\frac{dx}{d\tau })^{2}+V(x).
\end{equation}%
The Euclidian Feynman propagator can be evaluated in terms of
stationary-phase perturbation-method, in which the zero-order perturbation
comes from the action of classical trajectory of pseudo-particles in the
barrier region called the instantons. The explicit instanton solution of the
classical equation of motion
\[
m\frac{d^{2}x}{d\tau ^{2}}=\frac{dV(x)}{dx},
\]%
is found in our case as
\begin{equation}
x_{1}(\tau )=tn^{-1}\left( \frac{2\eta \tanh \left( \frac{\omega \tau }{2}%
\right) }{1-\eta ^{2}\tanh ^{2}\left( \frac{\omega \tau }{2}\right) }\right)
,
\end{equation}%
\begin{equation}
x_{2}(\tau )=tn^{-1}\left( \frac{2\eta ^{-1}\tanh \left( \frac{\omega \tau }{%
2}\right) }{1-\eta ^{-2}\tanh ^{2}\left( \frac{\omega \tau }{2}\right) }%
\right) ,
\end{equation}%
which exist in the large (counterclockwise rotation) and small (clockwise
rotation) barriers respectively (see Fig.3), where $tn(x)=sn(x)/cn(x)$ is
the Jacobian elliptic function, $\eta =\sqrt{(1-\mu )/(1+\mu )}$, and
\begin{equation}
\omega =\sqrt{4K_{1}(a-\lambda ^{2}V_{\min })}
\end{equation}%
is the oscillation frequency at the bottom of the potential well. Taking
into account of contributions of both instantons $x_{1}(\tau )$ and $%
x_{2}(\tau )$ with the corresponding boundary conditions Eq.(17) the Feynman
propagator is obtained up to the one-loop approximation and then the ground
state tunnel splitting is abstracted as
\begin{equation}
\Delta \varepsilon _{0}=Qe^{-\rho }\sqrt{2\left( \cosh \gamma +\cos 2\pi
s\right) }
\end{equation}%
where
\[
Q=\frac{2^{5/2}(a-\lambda ^{2}V_{\min })^{3/4}}{\sqrt{\left[ 1-\lambda
^{2}\left( 1-\mu ^{2}\right) \right] \pi }},
\]%
\[
\rho =\sqrt{\frac{a-\lambda ^{2}V_{\min }}{K_{1}\lambda ^{2}}}\ln \frac{%
1+\lambda \sqrt{1-\mu ^{2}}}{1-\lambda \sqrt{1-\mu ^{2}}},
\]%
and
\[
\gamma =\frac{2\mu }{\lambda ^{^{\prime }}}\sqrt{\frac{a-\lambda ^{2}V_{\min
}}{K_{1}}}\arctan \frac{\lambda ^{^{\prime }}\sqrt{1-\mu ^{2}}}{\mu }.
\]%
To see the particle-number parity effect clearly we consider the case of
vanishing Josephson coupling constant $J=0$ and thus $\mu =\gamma =0$. For
even-number of atoms ($s$ is integer) the tunnel splitting is
\[
\Delta \varepsilon _{0}=2Qe^{-\rho },
\]
however the tunnel splitting vanishes for odd-number of atoms ($s$
is half-integer) called the quench of quantum tunneling due to the
quantum phase interference of tunnel paths with the antiperiodic
boundary condition of wave function Eq.(17). Thus the degeneracy of
$\pm\frac{\pi}{2}$-phase states cannot be removed by quantum
tunneling. The tunnel splitting as a function of $J$ is shown in
Fig.6 for odd- (a) and even-number (b) $N$ respectively.

\section{conclusion}

We conclude that the BH Hamiltonian in the superstrong interaction regime
ought to be extended to include the two-body interaction of nearest
neighbors, which results in a fundamental phenomenon of many-body system,
namely the atom-pair tunneling. New dynamics of various oscillations
depending on the competition between the Josephson coupling constant $J$ and
interaction constant $U_{0}$ is found in superstrong interaction regime,
where the BH Hamiltonian gives rise to the fixed atom-occupation-number
state only. The QPT and the critical transition point are analyzed
analytically in terms of the potential-field method, which allows us to
convert the system of $N$ cold-atoms in a double-well trap to an effective
single-particle in a quasi-periodic potential in the elliptic integral
coordinate. The new oscillation states and related QPT should be observed in
practical experiment.

\begin{acknowledgements}
This work was supported by National Nature Science Foundation of China(Grant
No. 10775091).
\end{acknowledgements}

\appendix
\section
\bigskip The two-body interaction Hamiltonian in the second-quantization
formulation is%
\begin{equation}
H_{int}=\frac{1}{2}\int \Psi ^{\dag }(x_{2})\Psi ^{\dag
}(x_{1})U(|x_{1}-x_{2}|)\Psi (x_{1})\Psi (x_{2})dx_{1}dx_{2},
\label{A1}
\end{equation}%
with $\Psi (x)$\ being the field operator of ultracold atom-gas
clouds, which in the lattice-mode expansion\cite{liang,para} such
that
\begin{equation}
\Psi (x)=\sum_{i}^{N_{L}}a_{i}w_{i}(x)  \label{A2}
\end{equation}%
becomes%
\begin{equation}
H_{int}=\frac{1}{2}\sum\limits_{i,j,k,l}^{N_{L}}a_{i}^{\dag
}a_{j}^{\dag }a_{k}a_{l}\int
w_{i}(x_{2})w_{j}(x_{1})U(|x_{1}-x_{2}|)w_{k}(x_{1})w_{l}(x_{2})dx_{1}dx_{2},
\label{A3}
\end{equation}%
where four sums over the lattice sites are independent and $N_{L}$ denotes
the total number of lattice sites. In the strongly interacting regime the
sum over lattice sites should be extended to include nearest neighbors. The
arbitrary sum over lattice sites can be obtained up to the nearest-neighbor
approximation as%
\begin{equation}
H_{int}\approx H_{int-o}+H_{int-n}  \label{A4}
\end{equation}%
where the first term with all four modes on one-site
\begin{equation}
H_{int-o}=\frac{U_{0}}{2}\sum\limits_{i}^{N_{L}}a_{i}^{\dag
}a_{i}^{\dag
}a_{i}a_{i}=\frac{U_{0}}{2}\sum\limits_{i}^{N_{L}}n_{i}(n_{i}-1)
\label{A5}
\end{equation}%
is the well known on-site approximation in the BH Hamiltonian. The sum over
nearest-neighbor can be grouped under two configurations:
\[
H_{int-n}=H_{int-n}(2,2)+H_{int-n}(1,3),
\]%
where $H_{int-n}(2,2)$ denotes any two-mode on one-site and $H_{int-n}(1,3)$
denotes three-mode on one-site. Obviously
\begin{equation}
H_{int-n}(2,2)=(U_{1}+U_{2})\sum\limits_{i}^{N_{L}}n_{i}n_{i+1}+\frac{U_{2}}{%
2}\sum\limits_{i}^{N_{L}}(a_{i}^{\dag }a_{i}^{\dag
}a_{i+1}a_{i+1}+a_{i+1}^{\dag }a_{i+1}^{\dag }a_{i}a_{i})  \label{A6}
\end{equation}%
and%
\begin{eqnarray}
H_{int-n}(1,3) &=&\frac{U_{3}}{2}\sum_{i}^{N_{L}}(a_{i}^{\dag }a_{i+1}^{\dag
}a_{i+1}a_{i+1}+a_{i+1}^{\dag }a_{i}^{\dag }a_{i+1}a_{i+1}+a_{i+1}^{\dag
}a_{i+1}^{\dag }a_{i}a_{i+1}+a_{i+1}^{\dag }a_{i+1}^{\dag }a_{i+1}a_{i}
\nonumber \\
&&+a_{i+1}^{\dag }a_{i}^{\dag }a_{i}a_{i}+a_{i}^{\dag }a_{i+1}^{\dag
}a_{i}a_{i}+a_{i}^{\dag }a_{i}^{\dag }a_{i+1}a_{i}+a_{i}^{\dag }a_{i}^{\dag
}a_{i}a_{i+1})  \nonumber \\
&=&U_{3}\sum_{i}^{N_{L}}(n_{i+1}+n_{i}-1)(a_{i}^{\dag }a_{i+1}+a_{i+1}^{\dag
}a_{i}),  \label{A7}
\end{eqnarray}%
which can be combined with the hopping term in the Hamiltonian Eq.(1).

\bigskip

\bigskip

Figure Caption:

Fig.1 (color online)The time-evolution of average position for two-atom
occupation in the weakly ($J/U_{0}=1.5$) (a) and strongly ($J/U_{0}=0.2$, $%
0.1$) (b,c) interacting regime. Black dots denote the experimental data, red
solid-line is the value evaluated from the proposed Hamiltonian and blue
dash-line shows the result from Bose-Hubbard Hamiltonian.

Fig.2 (color online)Average position (a), visibility (b) and phase (c) for
single-atom occupation in strong interaction regime ($J/U_{0}=0.2$) with
black dots denoting the experimental data. Coinciding red (solid) and blue
(dash) lines are the values evaluated from the proposed and Bose-Hubbard
Hamiltonians respectively.

Fig.3 Effective potential in elliptic-integral coordinate $x$ (unit of $%
\mathcal{K}$) with asymmetric twin barriers and degenerate minima located at
$x_{\pm \text{ }}$. $x_{1}$ and $x_{2}$ denote the clockwise and
counterclockwise tunnel paths.

Fig.4 (color online)Phase-plane portraits for $\mu =0$ (a), $1>\mu >0$ (b),
and $\mu =1$ (c) and the variation of potential with respect to $\mu $.

Fig.5 (color online)The energy gap in unit of $K_{1}$ as a function of $J$
evaluated by the numerical diagonalization of the Hamiltonian Eq.(8) with $%
\Delta =0$ for the particle number $N=20$ and various values of $K_{2}$.

Fig.6 (color online)The energy gap as a function of $J$ evaluated by the
numerical diagonalization of the Hamiltonian for the particle number $N=7$
(a) and $N=8 $ (b).


\bigskip

\bigskip
\begin{references}
\bibitem{greiner}  M. Greiner, O. Mandel, T. Esslinger, T.W. H\={a}nsch,
and  I. Bloch, Nature {\bf 415}, 39 (2002).

\bibitem{spielman}  I.B. Spielman, W.D. Phillips, and J.V. Porto, Phys. Rev.
Lett. {\bf 98}, 080404 (2007).

\bibitem{stoferle}  T. St$\ddot{o}$ferle,  H. Moritz, C. Schori, M. K$%
\ddot{o}$hl,  and T. Esslinger, Phys. Rev. Lett. {\bf 92}, 130403
(2004).

\bibitem{albiez}  M. Albiez, R. Gati, J. F$\ddot{o}$lling, S. Hunsmann, M.
Cristiani, and M. K. Oberthaler, Phys. Rev. Lett. {\bf 95}, 010402
(2005).

\bibitem{folling}  S. F$\ddot{o}$lling, S. Trotzky, P. Cheinet, M. Feld, R.
Saers, A. Widera, T. M\={u}ller, and I. Bloch, Nature, {\bf 448},
1029 (2007).

\bibitem{zollner}  S. Z$\ddot{o}$llner, H.-D. Meyer, and  P. Schmelcher,
Phys. Rev. Lett. {\bf 100}, 040401 (2008).

\bibitem{de}  S. De Franceschi {\it et al}., Phys. Rev. Lett. {\bf 86}, 878
(2001).

\bibitem{zum}  D. M. Zumb\={u}'hl, C. M. Marcus, M. P. Hanson,  and  A. C.
Gossard, Phys. Rev. Lett. {\bf 93}, 256801 (2004).

\bibitem{duan}  L.-M. Duan, E. Demler, and M. D. Lukin, Phys. Rev. Lett. {\bf 91%
}, 090402 (2003).

\bibitem{kuklov}  A. B. Kuklov and B. V. Svistunov, Phys. Rev. Lett. {\bf 90},
100401 (2003).

\bibitem{lipkin} H.J. Lipkin, N. Meshkov, and N. Glick, Nucl. Phys. A {\bf 62}, 188 (1965).

\bibitem{liang}  J.-J. Liang, J.-Q. Liang, and W.-M. Liu, Phys. Rev. A {\bf %
68}, 043605, (2003).

\bibitem{para}  Gh.-S. Paraoanu, Phys. Rev. A {\bf 67}, 023607, (2003).

\end{references}
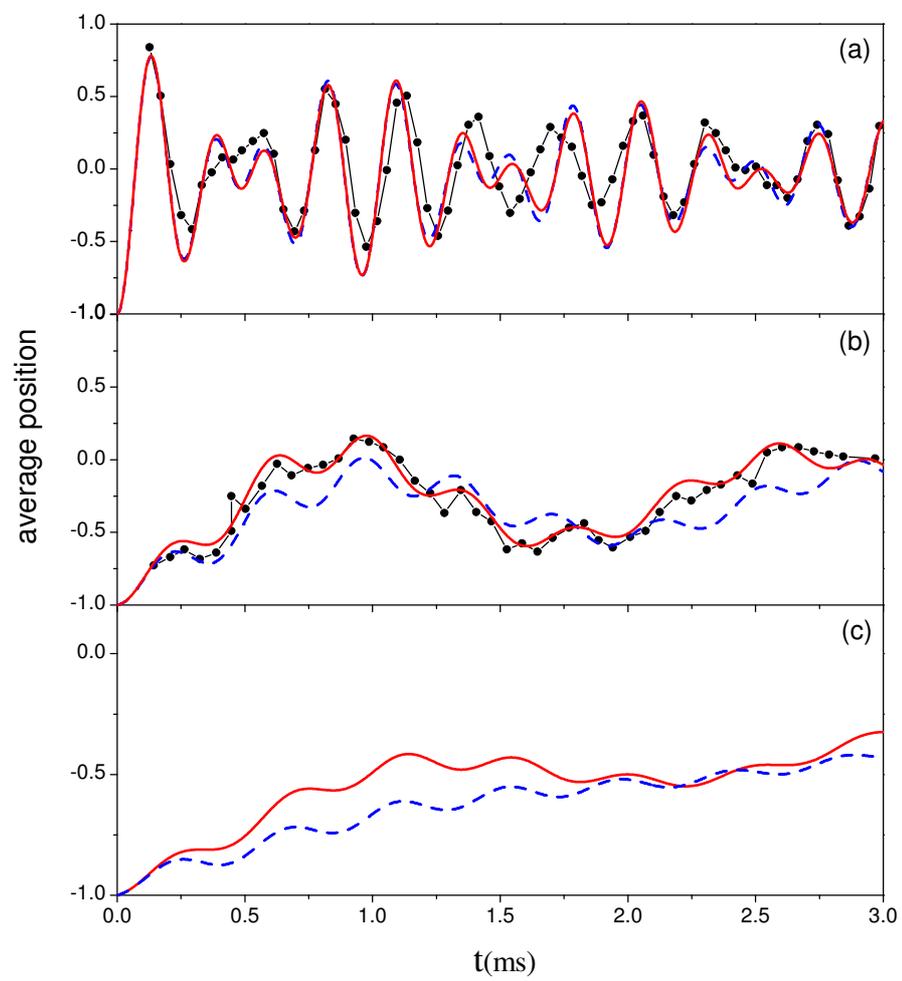
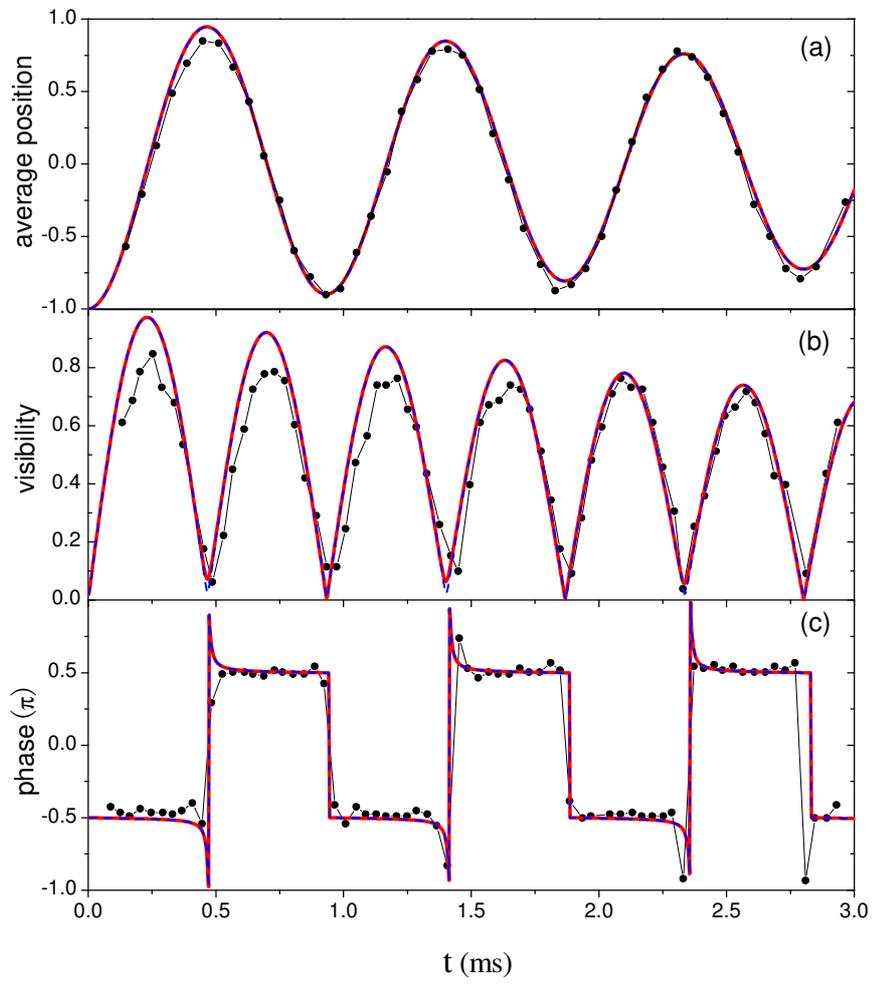
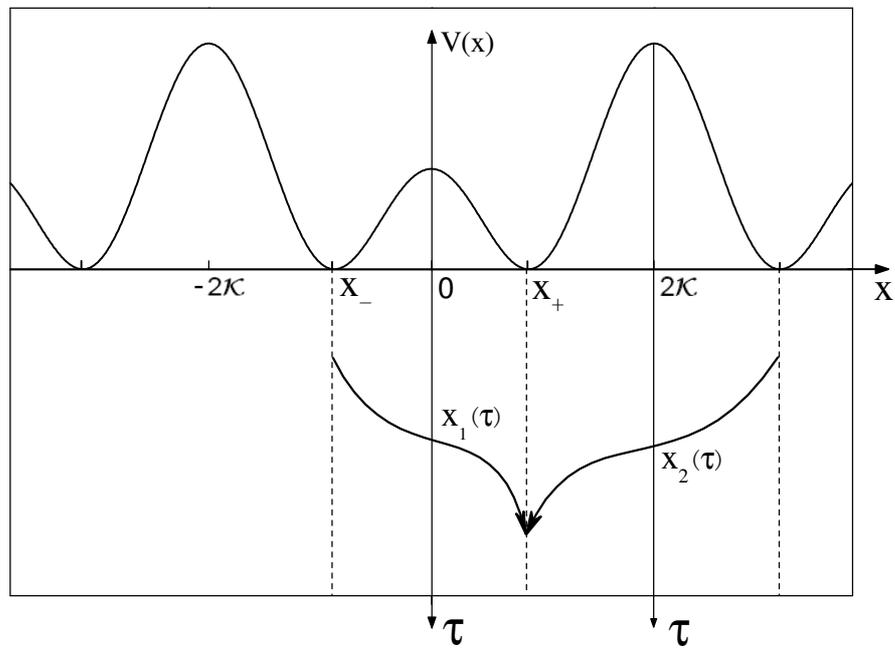
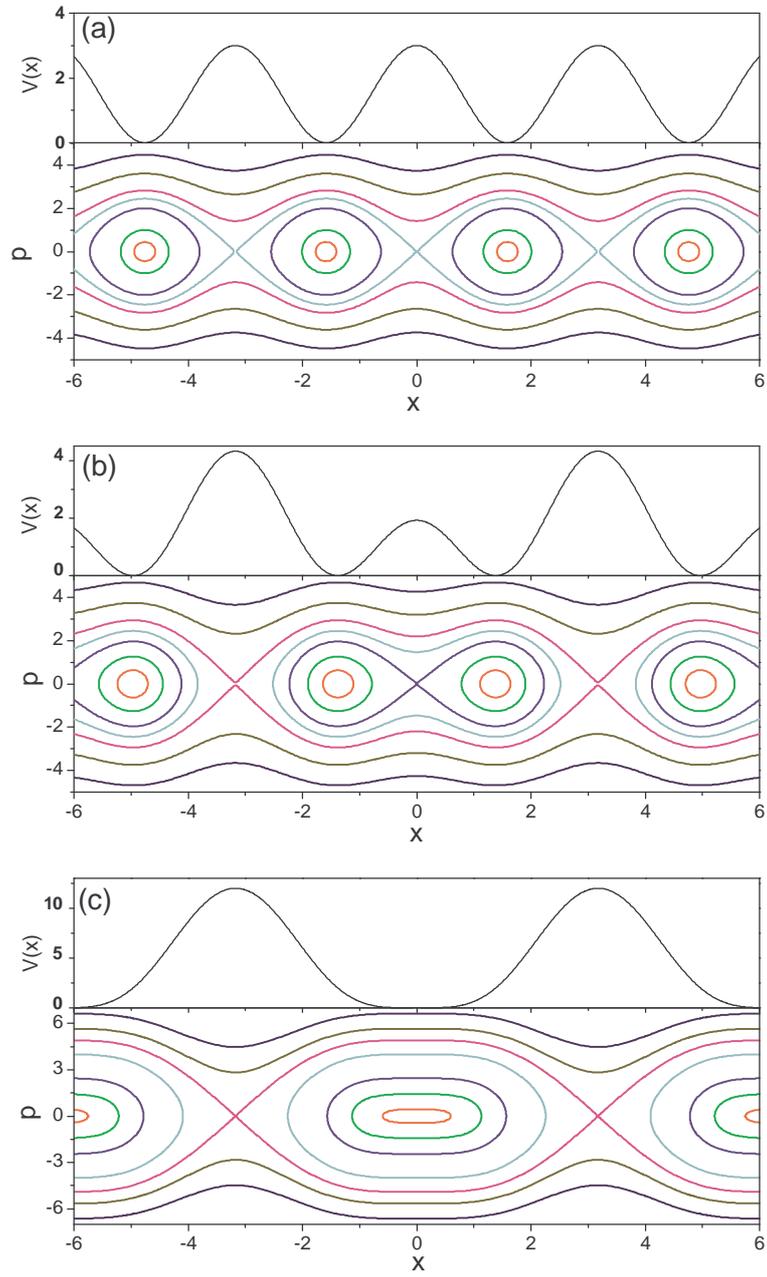
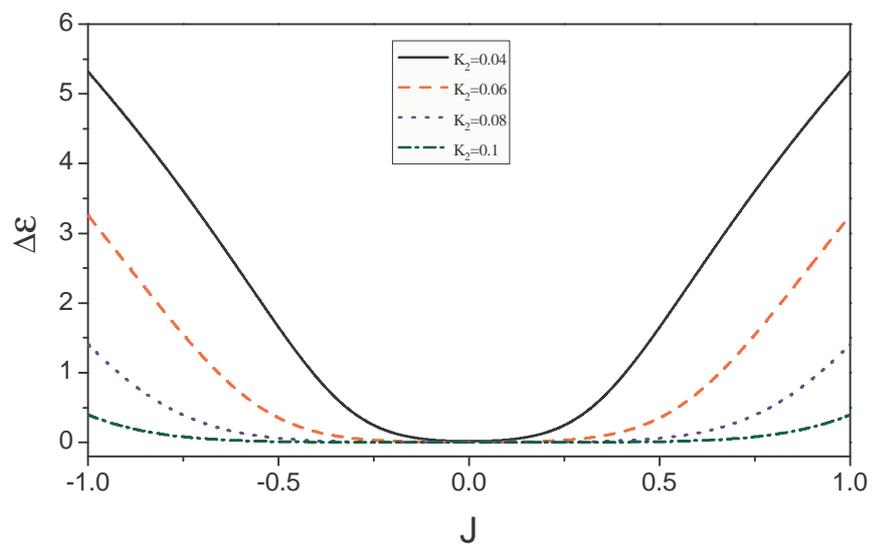
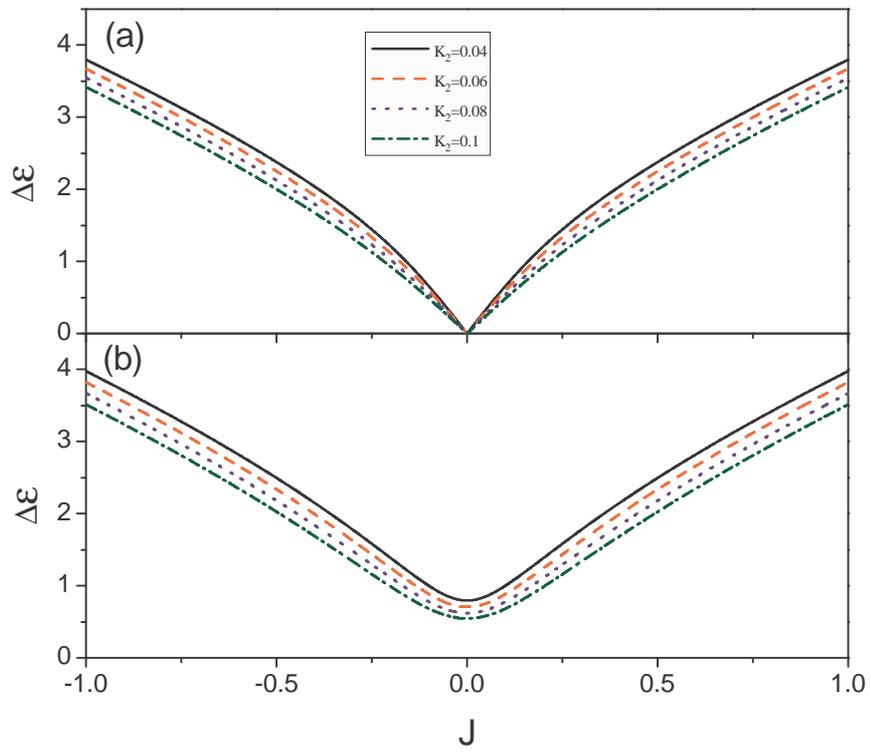
\end{document}